\documentclass[12pt, a4paper]{article}
\usepackage[utf8]{inputenc}
\usepackage[T2A]{fontenc}
\usepackage[english]{babel}

\usepackage{feynmf}
\usepackage{tabularx}
\usepackage{booktabs}
\usepackage{footnote}

\usepackage{indentfirst}
\usepackage{amsmath, amssymb, mathrsfs}

\usepackage{graphicx}

\usepackage{float}

\usepackage{caption}
\DeclareCaptionLabelSeparator{Sep}{\bf{:} }
\usepackage{csquotes}

\usepackage{array}
\usepackage{multirow}
\usepackage{hhline}

\usepackage{csquotes}
\usepackage{textcase}
\usepackage{xcolor}

\usepackage{hyperref}
\hypersetup{colorlinks,urlcolor=blue,linkcolor=blue,citecolor=blue}

\usepackage{geometry}
\usepackage{setspace}
\newcommand{\ETg}{\ensuremath{E_\text{T}^\gamma}}
\newcommand{\ETmiss}{\ensuremath{E_\text{T}^\text{miss}}}
\newcommand{\pTmissvec}{\ensuremath{\vec{p}_\text{T}^\text{~miss}}}
\newcommand{\ftzero}{\ensuremath{f_\text{T0}}/\Lambda^4}
\newcommand{\ftfive}{\ensuremath{f_\text{T5}}/\Lambda^4}
\newcommand{\fmzero}{\ensuremath{f_\text{M0}}/\Lambda^4}
\newcommand{\fmtwo}{\ensuremath{f_\text{M2}}/\Lambda^4}

\makeatletter
\def\maketitle{
\begin{center}\textbf{\Large{\@title}}
\par\vspace{2mm}\textbf{\@author}
\begin{singlespace}
\par$^a$National Research Nuclear University MEPhI, Moscow, Russia\par
*E-mail: AESemushin@mephi.ru\par
$^\dag$E-mail: EYSoldatov@mephi.ru
\end{singlespace}\end{center}
}
\makeatother

\title{Study of corrections for anomalous coupling limits due to the possible background BSM contributions}
\author{Artur E. Semushin$^{a,}$*, Evgeny Yu. Soldatov$^{a,\dag}$}
\date{}

\begin{document}
\maketitle
\begin{abstract}
    The search for anomalous couplings is one of the possible ways to find any deviations from the Standard Model. Effective field theory is used to parameterize the anomalous couplings in the Lagrangian with  operators of higher dimensions constructed from the SM fields. In the classical way, the limits on the Wilson coefficients of these operators are set based on beyond the Standard Model contributions induced for the signal process, whereas the ones induced for background processes are assumed to be negligible. This article provides a study of the corrections to the limits on Wilson coefficients by considering beyond the Standard Model contributions induced for background processes. The studies of $Z(\nu\bar{\nu})\gamma jj$ and $W(\ell\nu)\gamma jj$ productions in $pp$ collisions with $\sqrt{s}=13$~TeV and conditions of the ATLAS experiment at the LHC are used as an example. Cases with integrated luminosity collected during Run II of 139~fb$^{-1}$ and expected from Run III of 300~fb$^{-1}$ are considered. The expected 95\% CL limits on coefficients $f_\text{T0}/\Lambda^4$, $f_\text{T5}/\Lambda^4$, $f_\text{M0}/\Lambda^4$ and $f_\text{M2}/\Lambda^4$ are obtained both in the classical way and in the way, where the corrections from background anomalous contributions are applied. Corrected one-dimensional limits from $Z(\nu\bar{\nu})\gamma jj$ and $W(\ell\nu)\gamma jj$ productions are up to 9.1\% and 4.4\% (depending on the operator) tighter than the classical ones, respectively. Corrected combined limits are up to 3.0\% (depending on the operator) tighter than the classical ones. Corrections to two-dimensional limits are also obtained. The corrected contours are more stringent compared with the classical ones, and the maximal improvement is  17.2\%.
    \end{abstract}

\newpage
\section{Introduction}

In the Standard Model (SM), massive particles obtain their masses through a spontaneous symmetry breaking mechanism~\cite{Higgs1964,Englert1964,Guralnik1964}. A corresponding scalar particle, a Higgs boson, was discovered~\cite{ATLASHiggsObs2012,CMSHiggsObs2012} at the Large Hadron Collider (LHC) in 2012. Given the fact that the construction of the SM was almost finished at that stage, this theory still fails to explain important experimental and theoretical facts, such as the non-zero neutrino mass, existence of dark matter and dark energy, a large number of free parameters and their hierarchy in the theory. These gaps in the description indicate that the SM could be extended to a more general theory, explaining all aforementioned facts and deviations remain to be found. One can search for beyond-the-Standard-Model (BSM) effects (so-called new physics) that cause the deviations from the SM in two ways: direct and indirect. This means the direct search for new particles in the former case, and tests of the SM by searching for deviations in the interactions of the already known particles in the latter case. Since the direct search in LHC experiments gave no results~\cite{RappoccioDirectExoticReview,HinzmannSearchesForExotica}, the perspectives of the indirect search grew. The extensive search of deviations from the SM is crucial for the choice of a direction for the SM extension among a huge number of existing theoretical models.

The approach of an effective field theory (EFT)~\cite{DegrandeModernApproach} for the indirect search of new physics is used in this work. This effectively adds the new couplings induced by BSM physics to the existing theory. These additional couplings represent currently unavailable high energy effects of new physics.
Collision data from LHC experiments provides a possibility to set the limits on these new coupling constants, that is Wilson coefficients. The classical method for setting this kind of limits is based on considering the BSM contributions only in the main (signal) process. However, in the general case, one or several background processes can also have a BSM part, rising according to the probed coefficient  and contributing to the BSM yields. Changes in the BSM yields lead to  corrections in the derived limits~\cite{PoSPANIC2021}.

Vector boson scattering (VBS) $Z\gamma$ and $W\gamma$ productions are  processes that are  sensitive to  anomalous quartic gauge couplings (aQGC). Such processes, existing in the SM due to the non-abelian symmetry of the electroweak interaction, are very rare. Therefore, their study can lead to precise tests of the SM and the spontaneous symmetry breaking mechanism as well as searching for new physics. The choice of the neutrino decay of a $Z$ boson in this study provides the highest sensitivity, since its branching ratio is significantly larger than the charged-leptonic decay one. In addition, its final state identification efficiency is much higher than the hadronic decay one. The choice of the leptonic decay of a $W$ boson provides a large sensitivity due to its better identification efficiency compared with the hadronic decay one. VBS $Z\gamma$ and $W\gamma$ productions in $pp$ collisions require two associated hadronic jets at the tree level. $Z(\nu\bar{\nu})\gamma jj$ and $W(\ell\nu)\gamma jj$ are rare processes and were observed~\cite{ATLASZgammaExotRun2, CMSWgammaRun2} only recently in the analyses of collision data, collected during the full Run II of the LHC. This article uses these processes as example signals for considering the corrections to the limits on Wilson coefficients and to show a possible gain from this kind of correction.

\section{EFT framework}
EFT is a model-independent approach based on the parameterization of the Lagrangian with higher-dimensional operators in the following way:
\begin{equation}
\mathcal{L}=\mathcal{L}_\text{SM}+\sum\limits_{d>4}\sum\limits_i \frac{f_i^{(d)}}{\Lambda^{d-4}} \mathcal{O}_i^{(d)}, \label{EFTLagrangian}
\end{equation}
where $\mathcal{L}_\text{SM}$ denotes the SM Lagrangian (which dimension equals four), $\mathcal{O}_i^{(d)}$ is the $i$-th operator of dimension $d$, and the corresponding Wilson coefficient consists of dimensionless coupling constant $f_i^{(d)}$ and the new physics energy scale $\Lambda$. Such operators are constructed from the SM fields. They can describe vertices, predicted by the SM, enhancing corresponding couplings, as well as add new possible vertices to the theory.

The lowest operator dimension, where electroweak quartic gauge couplings without triple gauge coupling counterparts are predicted, is eight. Therefore, dimension-eight
operators are used to study aQGCs~\cite{EboliClassifying}. These operators can be split into three groups by their construction: operators of scalar (S), mixed (M) and tensor (T) families. They contain only derivatives of the Higgs doublet in the first case, derivatives of the Higgs doublet mixed with the electroweak field strength tensors in the second case and only the field strength tensors in the last case. Operators of the S-family do not affect couplings with photons; thus, the limits on corresponding Wilson coefficients cannot be set from $Z\gamma$ or $W\gamma$ production studies. In this work, corrections to the limits are considered for the coefficients of the following operators:
\begin{equation}
\begin{array}{l@{\qquad}l}
\mathcal{O}_\text{T0}=\text{Tr}\,\left[\hat{W}_{\mu\nu}\hat{W}^{\mu\nu}\right]\text{Tr}\,\left[\hat{W}_{\alpha\beta}\hat{W}^{\alpha\beta}\right], & \mathcal{O}_\text{T5}=\text{Tr}\,\left[\hat{W}_{\mu\nu}\hat{W}^{\mu\nu}\right]B_{\alpha\beta}B^{\alpha\beta}, \qquad \\[6pt]
\mathcal{O}_\text{M0}=\text{Tr}\,\left[\hat{W}_{\mu\nu}\hat{W}^{\mu\nu}\right] \left[\left(D_\beta \Phi\right)^\dag D^\beta \Phi\right], & \mathcal{O}_\text{M2}=B_{\mu\nu}B^{\mu\nu} \left[\left(D_\beta \Phi\right)^\dag D^\beta \Phi\right],
\end{array}
\end{equation}
since these operators affect couplings that can produce both final states, $Z\gamma$ and $W\gamma$. For instance, the operators
\begin{equation}
\mathcal{O}_\text{T8}=B_{\mu\nu}B^{\mu\nu}B_{\alpha\beta}B^{\alpha\beta}, \qquad \mathcal{O}_\text{T9}=B_{\alpha\mu}B^{\mu\beta}B_{\beta\nu}B^{\nu\alpha}
\end{equation}
affect only neutral aQGCs, and thus $W\gamma$ production cannot be used for setting or correcting limits on the corresponding Wilson coefficients.

If the Lagrangian is parameterized as  seen in Equation~\eqref{EFTLagrangian}, a squared amplitude of a generic process can be written as
\begin{multline}
\left|\mathcal{A}\right|^2=\left|\mathcal{A}_\text{SM} + \sum\limits_{d>4} \sum\limits_i \frac{f_i^{(d)}}{\Lambda^{d-4}} \mathcal{A}_{\text{BSM,}i}^{(d)}\right|^2=\left|\mathcal{A}_\text{SM}\right|^2 + \sum\limits_{d>4} \sum\limits_i \frac{f_i^{(d)}}{\Lambda^{d-4}} 2\text{Re}\, \mathcal{A}_\text{SM}^\dag \mathcal{A}_{\text{BSM,}i}^{(d)} + \\ + \sum\limits_{d>4} \sum\limits_i \left(\frac{f_i^{(d)}}{\Lambda^{d-4}}\right)^2 \left|\mathcal{A}_{\text{BSM,}i}^{(d)}\right|^2 + \sum\limits_{d,d^\prime>4} \sum\limits_{i,j,i>j} \frac{f_i^{(d)}}{\Lambda^{d-4}} \frac{f_j^{(d^\prime)}}{\Lambda^{d^\prime-4}} 2\text{Re}\, \mathcal{A}_{\text{BSM,}i}^{(d)\,\dag} \mathcal{A}_{\text{BSM,}j}^{(d^\prime)}.
\end{multline}
The full amplitude $\mathcal{A}$ consists of the SM amplitude $\mathcal{A}_\text{SM}$ and the sum over all BSM amplitudes $\mathcal{A}_{\text{BSM,}i}^{(d)}$, predicted by operators $\mathcal{O}_i^{(d)}$. In this general case, the squared amplitude consists of four parts: the SM term, the sum over all interference (or linear) terms, the sum over all quadratic terms and the sum over all cross terms, which represent an interference between different BSM amplitudes. The SM and quadratic terms are always positive, whereas interference and cross terms can be either positive  or negative. The procedure of limit setting with parameterization by a large number of operators (or with a large number of corresponding coefficients, which are the fit parameters of interest) is complicated. Thus, usually only one or two coefficients of EFT operators are treated as  non-zero, while all others are set to zero. This allows setting one- and two-dimensional (1D and 2D) limits on Wilson coefficients.

If the Lagrangian is parameterized by one (two) dimension-eight
operator, then the squared amplitude contains the SM term, one (two) interference term, one (two) quadratic term and zero (one) cross terms:
\begin{align}
&\left|\mathcal{A}\right|^2=\left|\mathcal{A}_\text{SM}\right|^2 + \frac{f}{\Lambda^4} 2\text{Re}\, \mathcal{A}_\text{SM}^\dag \mathcal{A}_\text{BSM}+ \frac{f^2}{\Lambda^8} \left|\mathcal{A}_\text{BSM}\right|^2, \label{SqAmp1dim} \\
&\nonumber \left|\mathcal{A}\right|^2=\left|\mathcal{A}_\text{SM}\right|^2 + \frac{f_1}{\Lambda^4} 2\text{Re}\, \mathcal{A}_\text{SM}^\dag \mathcal{A}_{\text{BSM,}1} + \frac{f_2}{\Lambda^4} 2\text{Re}\, \mathcal{A}_\text{SM}^\dag \mathcal{A}_{\text{BSM,}2} + \\
&\hspace{4cm}+ \frac{f_1^2}{\Lambda^8} \left|\mathcal{A}_{\text{BSM,}1}\right|^2 + \frac{f_2^2}{\Lambda^8} \left|\mathcal{A}_{\text{BSM,}2}\right|^2 + \frac{f_1 f_2}{\Lambda^8} 2\text{Re}\, \mathcal{A}_{\text{BSM,}1}^\dag \mathcal{A}_{\text{BSM,}2}. \label{SqAmp2dim}
\end{align}
Similar decomposition can be obtained for predicted yields. Note that the terms $\propto \Lambda^{-8}$ should be suppressed comparing with the terms $\propto \Lambda^{-4}$. Thus, there is some exceeding of the accuracy, since the terms $\propto \Lambda^{-8}$ are not discarded in this work. However, this full model guarantees positivity of the squared amplitude for any value of Wilson coefficients. Moreover, using such expansion up to the terms $\propto \Lambda^{-8}$, one should respect interference terms of some operators of dimensions higher than eight. Such operators are not classified yet, and their contribution is assumed to be suppressed by the  dimensionless coupling constant.

\section{Corrections to the limits}
Traditionally, in order to set the limits on Wilson coefficients, decomposition described in Equations~\eqref{SqAmp1dim} and~\eqref{SqAmp2dim} is applied only for the main (signal) process, while all background processes are assumed to have the SM term only. However, one or several backgrounds can be also affected by non-zero EFT coefficients. This leads to the changes in the BSM predicted yields. In this section,  the coefficient is set to its experimental limit value. The effect of the corrections in the BSM yields on the derived limits can be estimated from the following condition. If
\begin{equation}
\left| N_\text{BSM,sig} + N_\text{BSM,bkg} \right| > \left| N_\text{BSM,sig} \right|, \label{ImprCond}
\end{equation}
then the limits become tighter and weaker otherwise. Here, $N_\text{BSM,sig}$ and $N_\text{BSM,bkg}$ are the classical BSM prediction (from the signal process only) and its correction from the background processes, respectively. In the case of parameterization by one operator (as in Equation~\eqref{SqAmp1dim}), they are the sum of the corresponding interference and quadratic terms:
\begin{align}
&N_\text{BSM,sig(bkg)} = \frac{f}{\Lambda^4} N_{1,\text{sig(bkg)}} + \frac{f^2}{\Lambda^8} N_{2,\text{sig(bkg)}}.
\end{align}

In the low-sensitivity regime, when the derived limits are  large and the quadratic term dominates, i.e., $(f^2/\Lambda^8) N_{2,\text{sig(bkg)}} \gg |(f/\Lambda^4) N_{1,\text{sig(bkg)}}|$, the interference term can be dropped. Therefore, the condition in Equation~\eqref{ImprCond} is satisfied for any value of Wilson coefficient, and the corrected limits are tighter than the classical ones.

Otherwise, in the high-sensitivity regime, when the derived limits are  small and the interference term dominates, i.e., $(f^2/\Lambda^8) N_{2,\text{sig(bkg)}} \ll |(f/\Lambda^4) N_{1,\text{sig(bkg)}}|$, the quadratic term can be dropped. Therefore, the condition in Equation~\eqref{ImprCond} turns into the condition $|N_{1,\text{sig}}+N_{1,\text{bkg}}| > |N_{1,\text{sig}}|$, and there are two possible cases, depending on a relative sign of $N_{1,\text{sig}}$ and $N_{1,\text{bkg}}$. In the first case, when $N_{1,\text{sig}} N_{1,\text{bkg}} > 0$, the limits become tighter. In the second case, when $N_{1,\text{sig}} N_{1,\text{bkg}} < 0$, the limits can become either tighter if $|N_{1,\text{bkg}}|>2 |N_{1,\text{sig}}|$ or weaker if $|N_{1,\text{bkg}}|<2 |N_{1,\text{sig}}|$.

In the medium-sensitivity regime, when the interference and quadratic terms are of the same order, it is possible that the limits become tighter from one side and weaker from another side. The study of this regime is complicated due to the large number of cases. Thus, in general case, the limit corrections can lead to either tightening (improvement) or weakening of the limits. The current sensitivity of leading high-energy experiments (e.g., Run II and Run III of the LHC) to aQGC is  low (but is close to the medium regime). In this case, the corrections to the limits lead to tightening of the limits from both sides.

\section{Simulation of signal and background events, event selection}
\begin{figure}[h!]
    \centering
    \includegraphics[width=\textwidth]{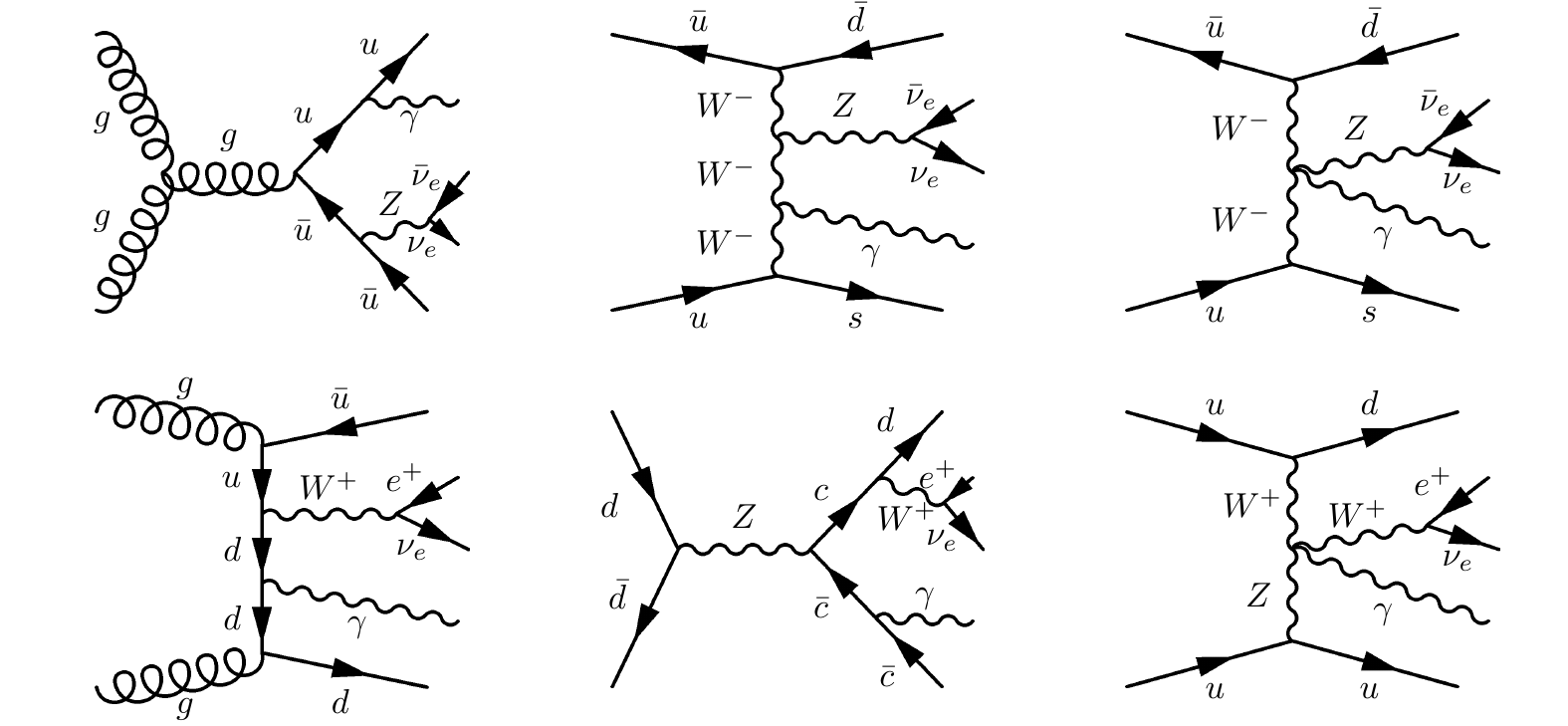}
    \caption{Examples of the Feynman diagrams for $Z(\nu\bar{\nu})\gamma$ (top) and $W(\ell\nu)\gamma$ (bottom) partially strong non-VBS (left), electroweak non-VBS (center) and VBS (right) productions in association with two~jets.}
    \label{Diagrams}
\end{figure}

The methodology for the correction of limits is applied to $Z\gamma$ and $W\gamma$ productions in association with two jets. As it was mentioned above the neutrino decay of a $Z$ boson and leptonic decay of a $W$ boson are considered to increase the sensitivity to aQGC. These processes are sensitive to aQGC since boson pairs can be produced via a VBS subprocess. In addition to the VBS mechanism, which has an electroweak nature in the SM,  diboson states with two jets can be also produced via non-VBS mechanisms: partially strong and fully electroweak  that have two and zero strong vertices at the tree level, respectively. Examples of the diagrams for VBS and non-VBS $Z(\nu\bar{\nu})\gamma$ and $W(\ell\nu)\gamma$ productions are presented in Figure~\ref{Diagrams}. The $Z\gamma$ boson pair can be produced through the VBS mechanism either in quartic gauge coupling with charged bosons ($WWZ\gamma$), predicted by the SM, or in neutral quartic gauge couplings ($ZZZ\gamma$, $ZZ\gamma\gamma$, $Z\gamma\gamma\gamma$), that are absent in the SM at the tree level. The $W\gamma$ boson pair can be produced through VBS in the quartic gauge couplings predicted by the SM ($WWZ\gamma$, $WW\gamma\gamma$), whose contribution to the process amplitude can be changed due to the dimension-eight
EFT operators.

All considered processes were generated with a Monte Carlo (MC) event generator \textsc{MadGraph5\_aMC@NLO}~\cite{MadGraph}  using $pp$ collisions with center-of-mass energy ($\sqrt{s}$) of 13~TeV. The capabilities of this generator were used for the  generation of different terms of squared amplitude as described in Equations~\eqref{SqAmp1dim} and~\eqref{SqAmp2dim}, separately~\cite{PHAN2021}. \textsc{Pythia8}~\cite{Pythia} was used
to model the parton shower, hadronization, and underlying
event. \textsc{Delphes3}~\cite{Delphes} with ATLAS detector geometry was used for detector simulation. The Run~II integrated luminosity of 139~fb$^{-1}$ and Run~III integrated luminosity approximate estimation of 300~fb$^{-1}$ were used for the calculations.

To study the corrections to the aQGC limits from $Z(\nu\bar{\nu})\gamma$ and $W(\ell\nu)\gamma$ productions, different phase-space regions are used in this work --- the $Z\gamma$ and $W\gamma$ regions, respectively. They correspond to the following signatures: one tight photon, large missing transverse energy $\ETmiss$, at least two hadronic jets and exactly zero (one) leptons for the $Z\gamma$ ($W\gamma$) region.
Backgrounds that can be also produced via anomalous quartic gauge couplings and, therefore, are used to demonstrate the corrections to the limits on Wilson coefficients, are $W(\ell\nu)\gamma$ and $Z(\ell\bar{\ell})\gamma$ in the $Z\gamma$ region, and $Z(\ell\bar{\ell})\gamma$ in the $W\gamma$ region. Additional background that is considered is $t\bar{t}\gamma$ production, which is significant in both regions and has no BSM part. The toy model in this work does not include backgrounds, which are usually estimated using data-driven techniques (e.g., backgrounds from particle misidentification, such as $e\rightarrow\gamma$, $j\rightarrow\gamma$ misidentification or fake missing transverse energy from jets energy mismeasurement), since they are not dominant  \cite{ATLASZgammaExotRun2, ATLASZgammaRun1, ATLASZnunugamma2022}, especially in the high-energy regions  where the EFT sensitivity is the highest.

The object reconstruction and selection used in this study are as close to ATLAS Run~I study of electroweak $Z(\nu\bar{\nu})\gamma$ production~\cite{ATLASZgammaRun1} ones as this is possible in \textsc{Delphes3}. The event selection criteria for the $Z\gamma$ and $W\gamma$ regions are based on the ones from the ATLAS Run~I study of electroweak $Z(\nu\bar{\nu})\gamma jj$ production~\cite{ATLASZgammaRun1} and from the CMS Run~II study of electroweak $W(\ell\nu)\gamma jj$ production~\cite{CMSWgammaRun2}, respectively. However, some ineffective cuts were excluded, and the summary of the selection is given in Table~\ref{Selection}.
\begin{table}[h!]
    \caption{Event selection criteria.}
    \label{Selection}
    \newcolumntype{C}{>{\centering\arraybackslash}X}
    \begin{tabularx}{\textwidth}{CC}
        \toprule
        $Z\gamma$ region & $W\gamma$ region \\ \midrule
        $\ETg > 150$ GeV & $\ETg > 100$ GeV \\
        $\ETmiss > 100$ GeV & $\ETmiss > 30$ GeV \\
        $\left| \Delta\varphi \left( \pTmissvec\textsuperscript{ 1} , \gamma jj \right) \right|  > 3\pi /4$ & $p_\text{T}^\ell > 30$ GeV \\
        $\left| \Delta\varphi \left( \pTmissvec , \gamma \right) \right| > \pi /2$ & $m_\text{T}^W$ \textsuperscript{3} $> 30$ GeV \\
        $\left| \Delta\varphi \left( \pTmissvec , j \right) \right| > 1$ & $\left| m_{e\gamma}-m_Z \right| > 10$ GeV \\
        $\Delta R$ \textsuperscript{2} $> 0.4$ for any objects pair & $\Delta R > 0.5$ for any objects pair \\
        & $\left| \Delta \eta_{jj} \right| >2.5$ \\
        \bottomrule
    \end{tabularx}
    \noindent{\footnotesize{
    \textsuperscript{1} $\pTmissvec$ is the missing transverse momentum vector, and missing transverse energy $\ETmiss$ is its absolute value. \\
    \textsuperscript{2} $\Delta R = \sqrt{(\Delta \eta)^2 + (\Delta \varphi)^2}$ is the angular distance between two objects, where $\eta = -\ln\tan (\theta/2)$ is the pseudorapidity. \\
    \textsuperscript{3} $m_\text{T}^W = \sqrt{2 p_\text{T}^\ell \ETmiss (1-\cos\Delta\varphi_{\ell,\pTmissvec})}$ is the transverse mass of the $W$ boson.}}
\end{table}

\begin{figure}[h!]
    \centering
    \begin{minipage}{0.49\textwidth}
    \centering
    \includegraphics[width=0.74\textwidth]{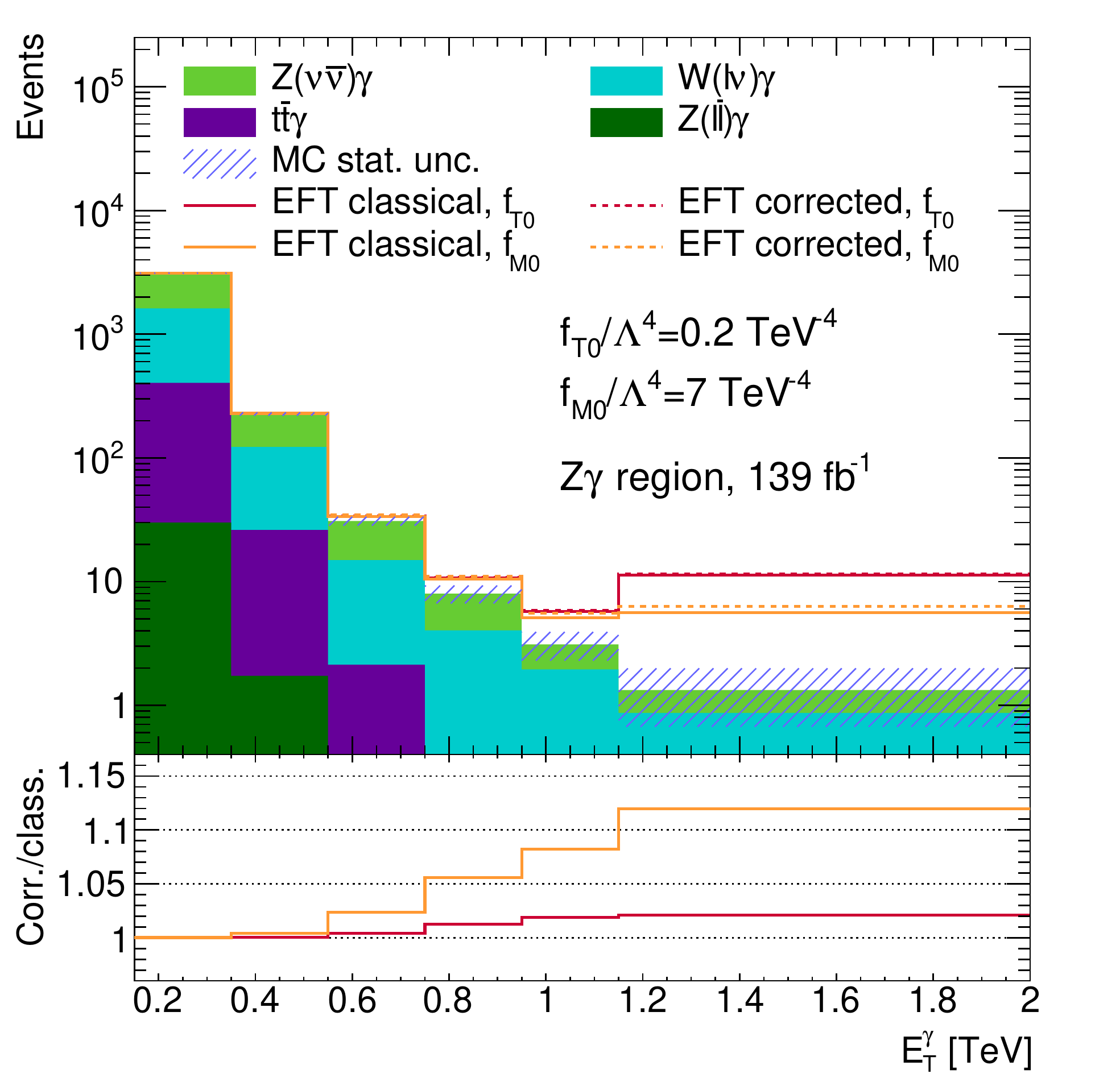}
    \end{minipage}
    \begin{minipage}{0.49\textwidth}
    \centering
    \includegraphics[width=0.74\textwidth]{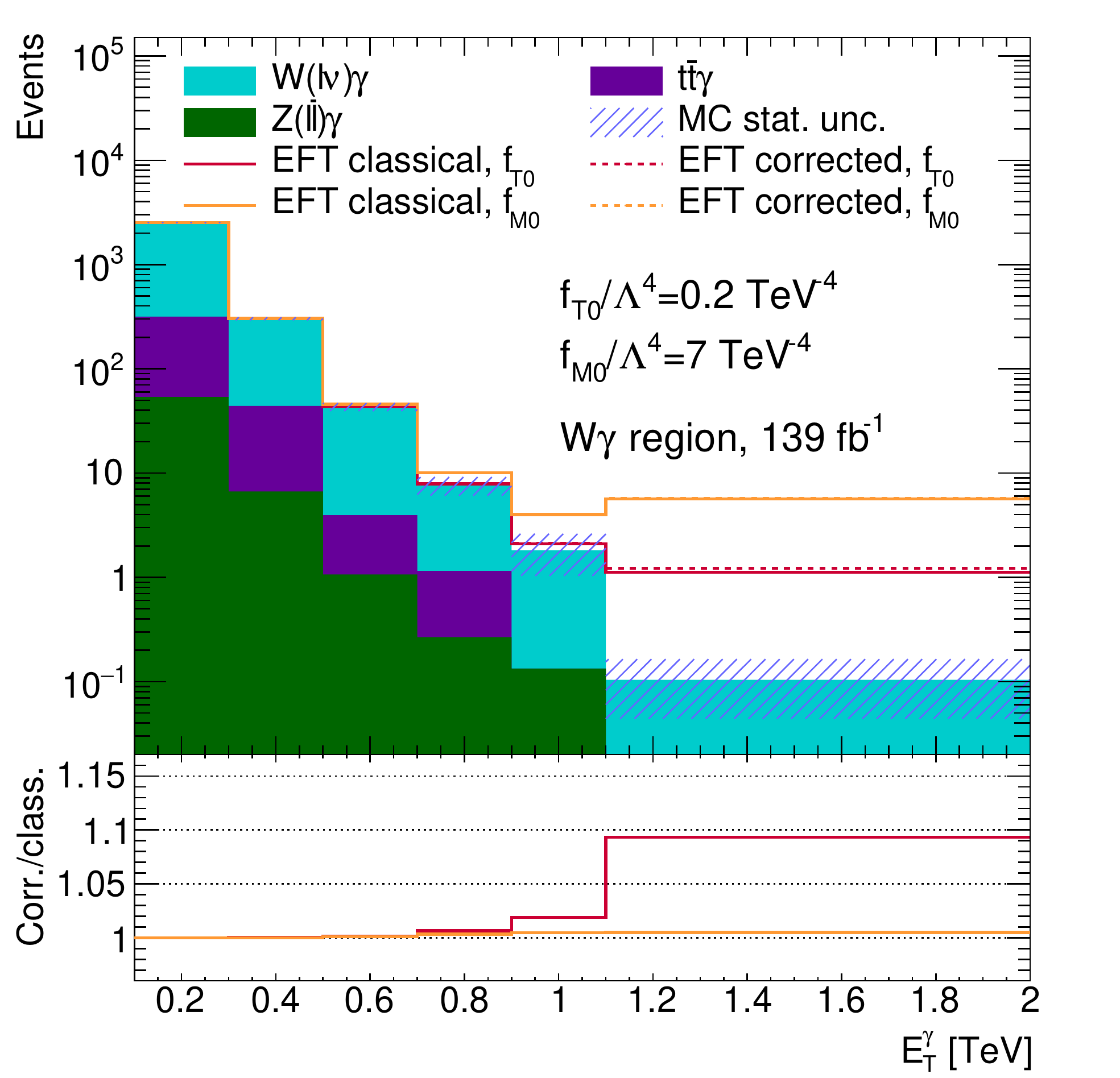}
    \end{minipage}
    \par
    \begin{minipage}{0.49\textwidth}
    \centering
    \includegraphics[width=0.74\textwidth]{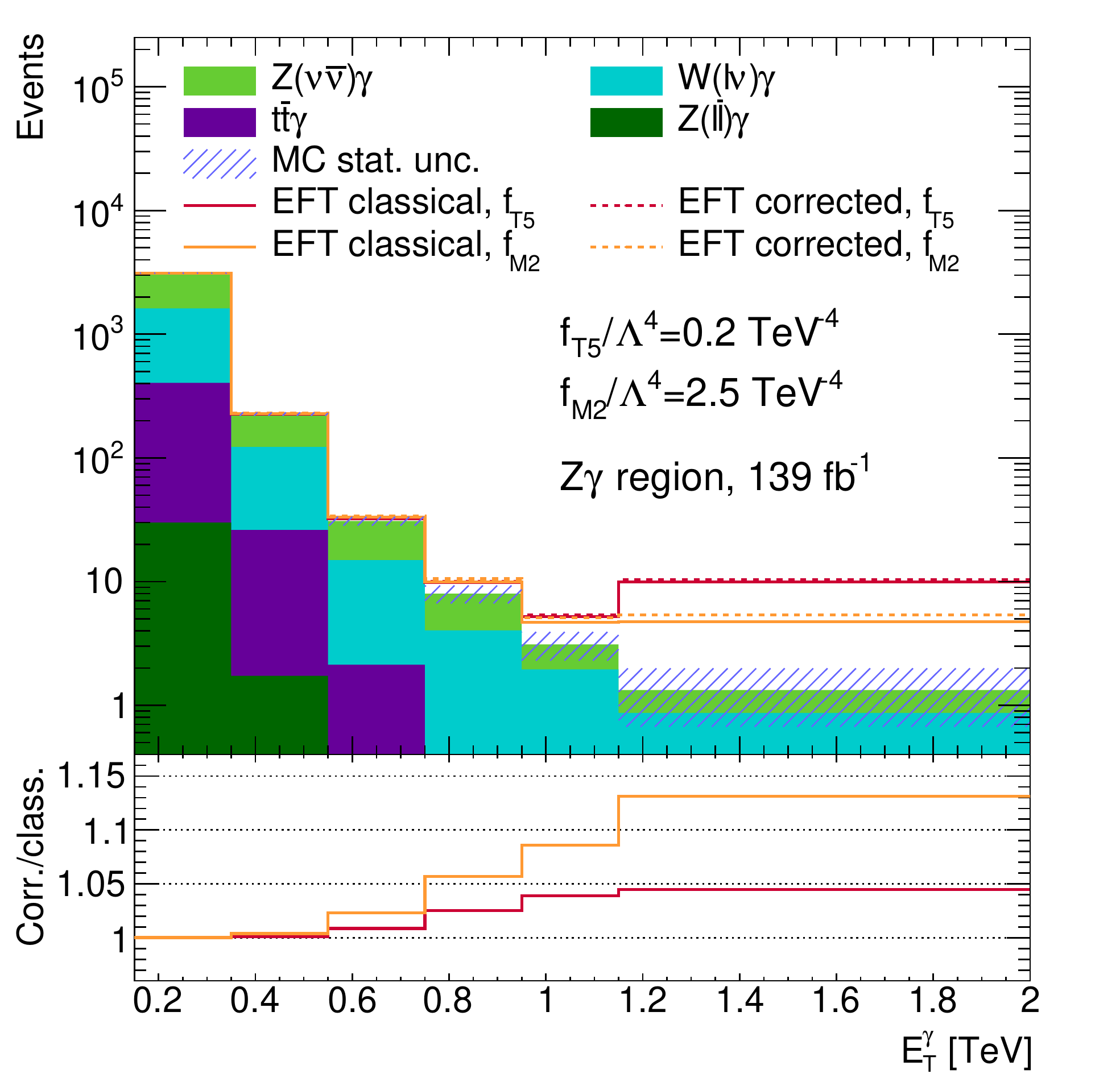}
    \end{minipage}
    \begin{minipage}{0.49\textwidth}
    \centering
    \includegraphics[width=0.74\textwidth]{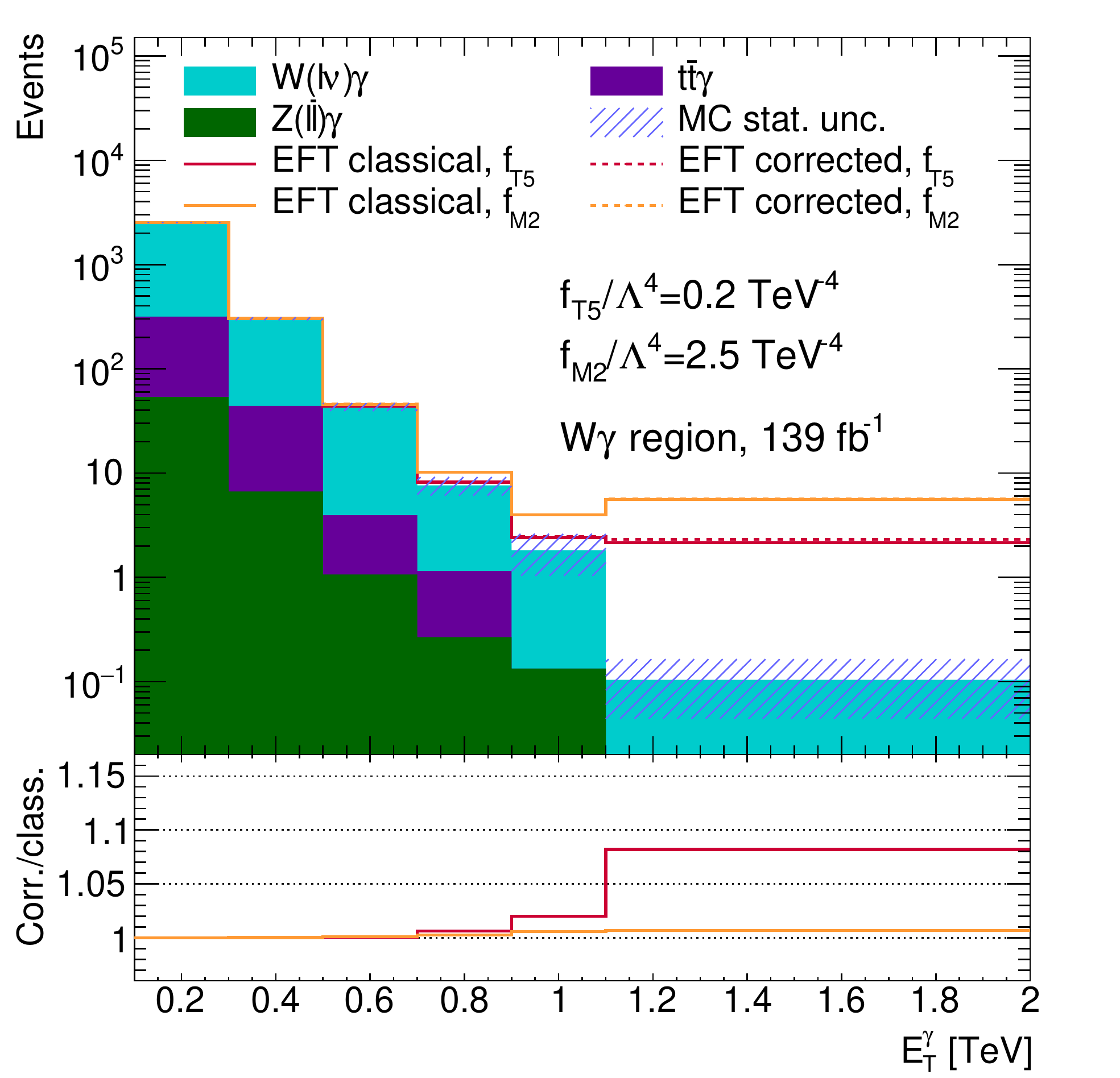}
    \end{minipage}
    \caption{$\ETg$ distributions in the $Z\gamma$ (left) and $W\gamma$ (right) regions for Run II integrated luminosity of 139~fb$^{-1}$. Solid (dashed) lines correspond to the classical (corrected) EFT prediction for cases if one of the considered coefficients is non-zero. The histogram in the the lower panel is the ratio of corrected and classical EFT predictions. The last bin includes all overflows.}
    \label{Distributions}
\end{figure}

The sensitivity to aQGC grows rapidly with increase of the center-of-mass energy of VBS. However, this energy cannot be measured due to the neutrino presence in the final state. The transverse energy of a photon $\ETg$ highly correlates with the center-of-mass energy of VBS. Thus, $\ETg$ distribution (Figure~\ref{Distributions}) is used for setting the limits in this study.

\section{Statistical treatment}
In this section, $\boldsymbol{\mu}$ corresponds to a vector in the parameter-of-interest space --- that is a single Wilson coefficient ($\boldsymbol{\mu}=f/\Lambda^4$) in the case of setting 1D limits or a couple of different Wilson coefficients ($\boldsymbol{\mu}=(f_1/\Lambda^4,f_2/\Lambda^4)$) in the case of setting 2D limits, and $\boldsymbol{\theta}$ corresponds to a vector of nuisance parameters, that represents the uncertainties. In the limit-setting procedure, the likelihood-based test statistic~\cite{CowanAsymptFormulae}
\begin{equation}
t_{\boldsymbol{\mu}}=-2\ln \frac{L(\boldsymbol{\mu},\hspace{0.15em}\hat{\hat{\hspace{-0.15em}\boldsymbol{\theta}}}(\boldsymbol{\mu}))}{L(\hat{\boldsymbol{\mu}},\hat{\boldsymbol{\theta}})}
\end{equation}
is used, where the denominator contains the likelihood function at its global maximum and the numerator contains the likelihood function, maximized by nuisance parameters for a given point $\boldsymbol{\mu}$ in the parameter-of-interest space. The likelihood contains the product of Poisson distributions of numbers of events in each bin of the $\ETg$ distribution from Figure~\ref{Distributions} and Gaussian constraint of the nuisance parameters:
\begin{equation}
L(\boldsymbol{\mu},\boldsymbol{\theta}) = \prod\limits_i \frac{(N^i_\text{pred}(\boldsymbol{\mu},\boldsymbol{\theta}))^{N^i_\text{data}}}{N^i_\text{data}!} e^{-N^i_\text{pred}(\boldsymbol{\mu},\boldsymbol{\theta})} \times \prod\limits_j \frac{1}{\sqrt{2\pi}} e^{-\theta_j^2/2}.
\end{equation}

This study does not use collision data events, and thus only expected limits are set; therefore, the expected number of signal and background events from SM is used as pseudodata $N^i_\text{data}$. The predicted number of events is parameterized by Wilson coefficients according to Equations~\eqref{SqAmp1dim} and~\eqref{SqAmp2dim} and by nuisance parameters as\vspace{-6pt}
\begin{align}
& N^i_\text{pred}(\boldsymbol{\mu},\boldsymbol{\theta}) = \left( N^i_0 (1+\sigma^i_{0}\theta^i_{0}) + \frac{f}{\Lambda^4} N^i_1 (1+\sigma^i_{1}\theta^i_{1}) + \frac{f^2}{\Lambda^8} N^i_2 (1+\sigma^i_{2}\theta^i_{2}) \right) (1+\sigma_\text{syst} \theta_\text{syst}), \\
&\nonumber N^i_\text{pred}(\boldsymbol{\mu},\boldsymbol{\theta}) = \left( N^i_0 (1+\sigma^i_0\theta^i_0) + \frac{f_1}{\Lambda^4} N^i_{1,1} (1+\sigma^i_{1,1}\theta^i_{1,1}) + \frac{f_2}{\Lambda^4} N^i_{1,2} (1+\sigma^i_{1,2}\theta^i_{1,2}) +\right. \\
&\hspace{0.5cm} \left.+ \frac{f_1^2}{\Lambda^8} N^i_{2,1} (1+\sigma^i_{2,1}\theta^i_{2,1}) + \frac{f_2^2}{\Lambda^8} N^i_{2,2} (1+\sigma^i_{2,2}\theta^i_{2,2}) + \frac{f_1 f_2}{\Lambda^8} N^i_{12} (1+\sigma^i_{12}\theta^i_{12}) \right) (1+\sigma_\text{syst} \theta_\text{syst})
\end{align}
in the case of setting 1D and 2D limits. Here, $\sigma$ is a variation, corresponding to an uncertainty represented by $\theta$. In this study, only one total systematic uncertainty of $\sigma_\text{syst}=0.2$ for all bins is used, since it is a conservative estimation of systematic uncertainties, which are mentioned in~\cite{ATLASZgammaRun1,CMSWgammaRun2}. Nuisance parameters corresponding to the statistical uncertainties are created for SM and each BSM term in each bin.

The CL$_{s+b}$ technique, used in this study, determines the 95\% CL confidence region as a region in the parameter-of-interest space, where the $p$-value
\begin{equation}
p_{\boldsymbol{\mu}} = \int\limits_{t_{\boldsymbol{\mu}}^\text{obs}}^{\infty} f(t_{\boldsymbol{\mu}} | \boldsymbol{\mu}) \, \text{d} t_{\boldsymbol{\mu}} > 0.05,
\end{equation}
$t_{\boldsymbol{\mu}}^\text{obs}$ is the observed value of the test statistic and $f(t_{\boldsymbol{\mu}}|\boldsymbol{\mu})$ is the test statistic distribution under the $\boldsymbol{\mu}$ hypothesis (i.e., the hypothesis that the data is distributed according to the $\boldsymbol{\mu}$ value). This distribution is assumed to be asymptotic, i.e., chi-square distribution with one or two degrees of freedom for 1D or 2D limits, due to the Wilks' theorem~\cite{Wilks}. Therefore, the procedure of setting 95\%~CL limits means to solve the equation $t_{\boldsymbol{\mu}}^\text{obs}=3.84\,(5.99)$ for 1D (2D) limits. It should be noted, that the modified statistical method CL$_s$ is also used in some analyses of LHC data. This method yields slightly less stringent limits, however the corrections to such limits are expected to be of the same order.

\section{Results}
The resulting 95\% CL expected classical and corrected limits on considered Wilson coefficients from $Z\gamma$ and $W\gamma$ regions can be found in Tables~\ref{Zgamma1DResults} and~\ref{Wgamma1DResults}. Moreover, these tables contain an additional column, which shows how much more stringent the corrected confidence interval is when compared to the classical one. In the $Z\gamma$ ($W\gamma$) region, the improvement is the highest for coefficients of operators of the M-family (T-family). This is because $W\gamma$ production has a larger (smaller) sensitivity to new physics, described by the considered M-family (T-family) operators, and vice versa for $Z\gamma$ production.

Despite the fact that the limits are usually obtained from the single process measurements, LHC experiments aim for combinations of individual channels into a global fit~\cite{ATLASCMSCombTGC2016, CMSWWandWZ2020, ATLASEFTComb2021, ATLASEFTCombHigVec2022}. Table~\ref{Combination1DResults} shows the limits set from a  combined region, i.e., the combination of the limits from the previous two tables. Such a combination leads to more stringent limits than ones from the separate regions. Combined region, and therefore corresponding likelihood, contains all bins from $\ETg$ distribution in the $Z\gamma$ and $W\gamma$ regions (Figure~\ref{Distributions}). Improvement of the limits from corrections in this region lies between the improvements in the  $Z\gamma$ and $W\gamma$ regions.

\begin{figure}[h!]
    \centering
    \begin{minipage}{0.32\textwidth}
    \centering
    \includegraphics[width=0.95\textwidth]{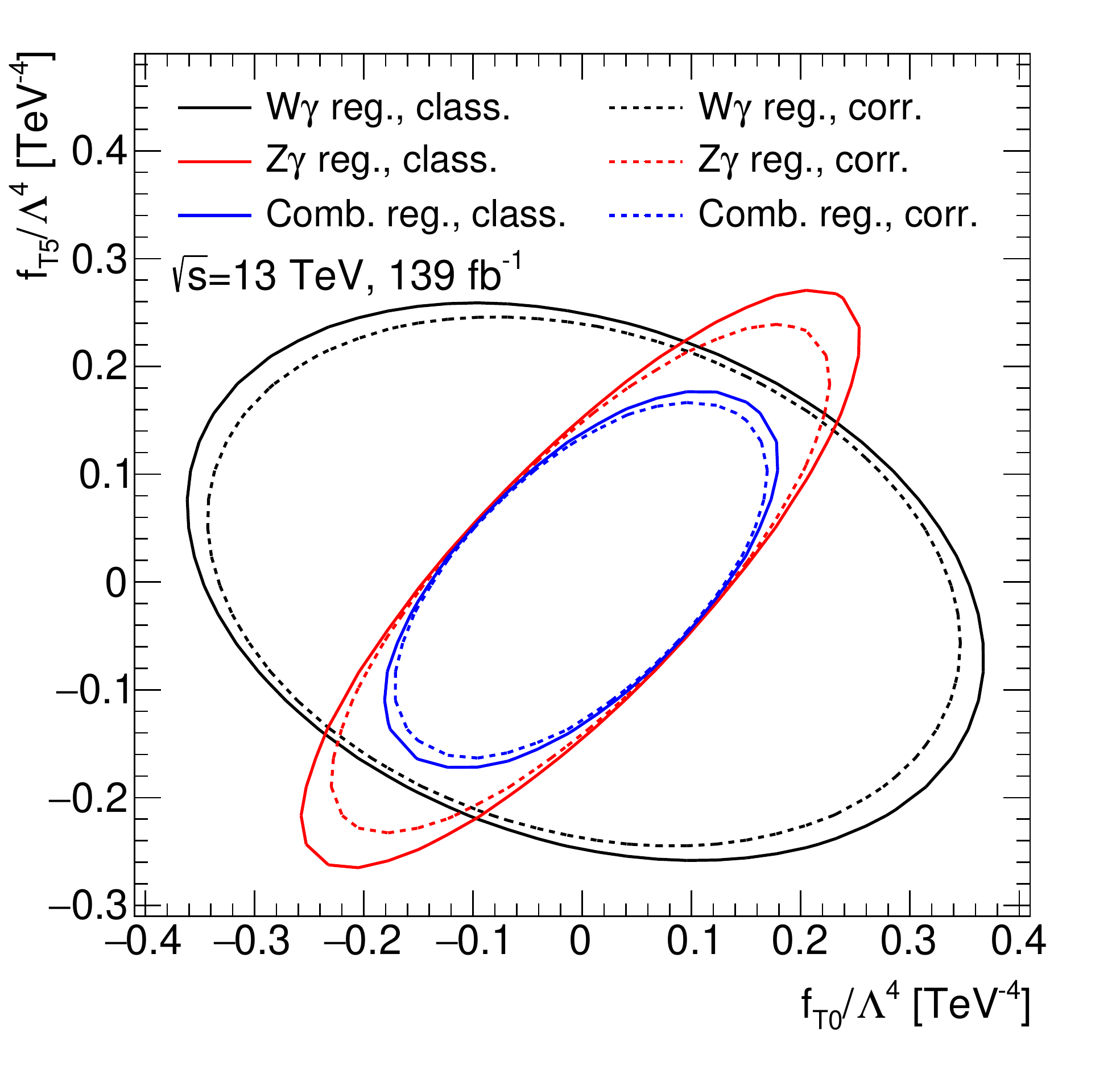}
    \end{minipage}
    \begin{minipage}{0.32\textwidth}
    \centering
    \includegraphics[width=0.95\textwidth]{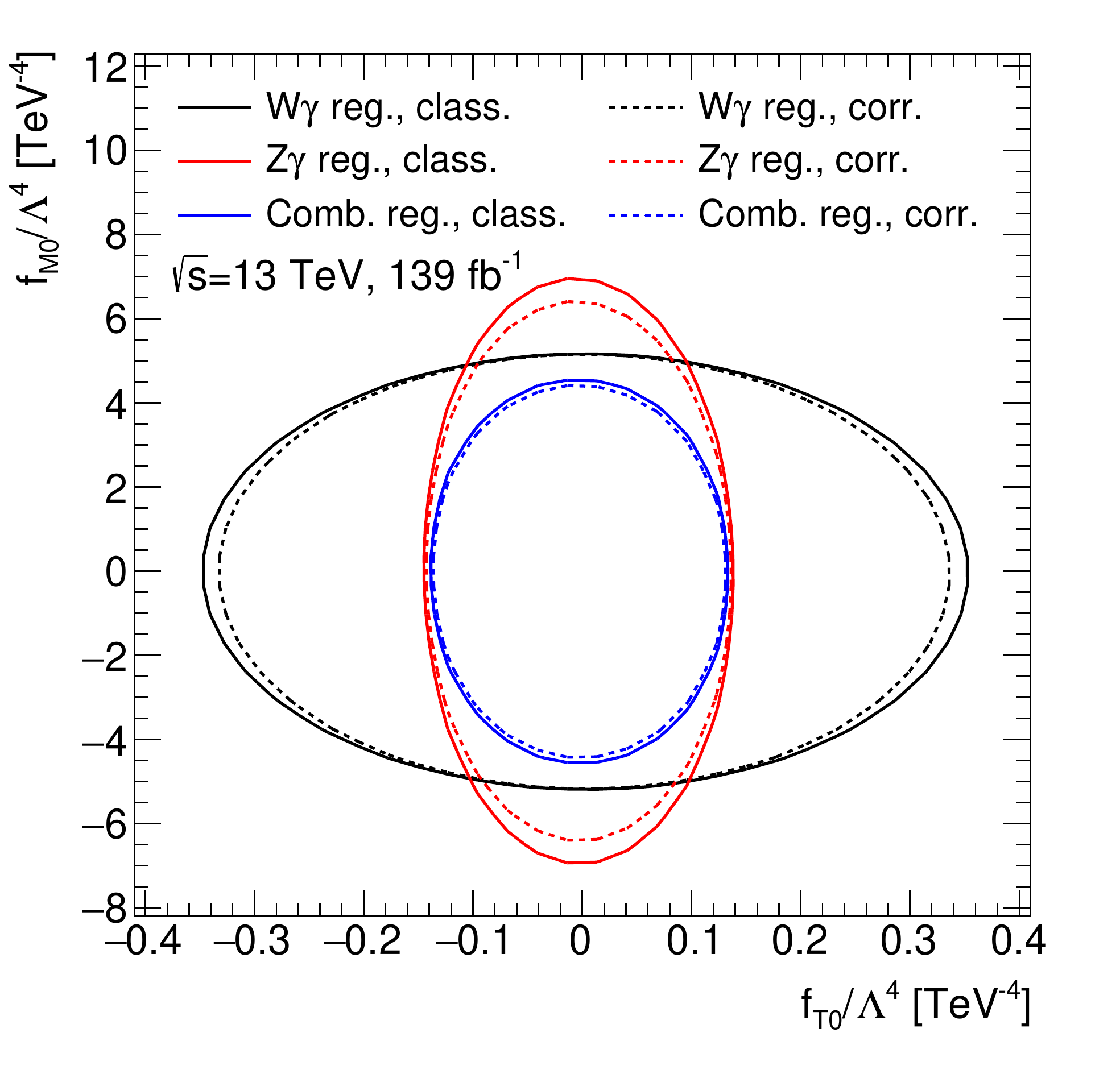}
    \end{minipage}
    \begin{minipage}{0.32\textwidth}
    \centering
    \includegraphics[width=0.95\textwidth]{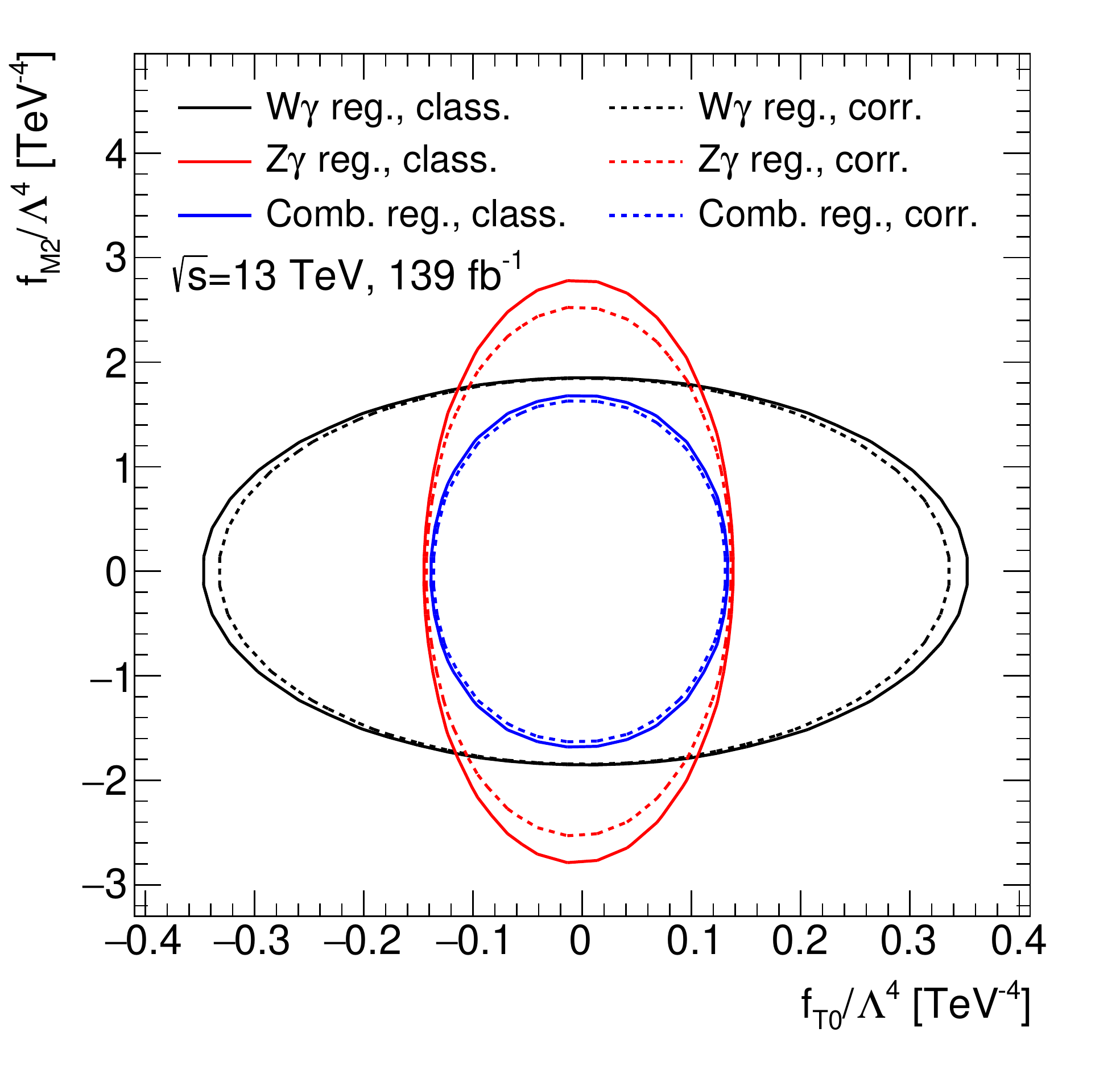}
    \end{minipage}
    \par
    \begin{minipage}{0.32\textwidth}
    \centering
    \includegraphics[width=0.95\textwidth]{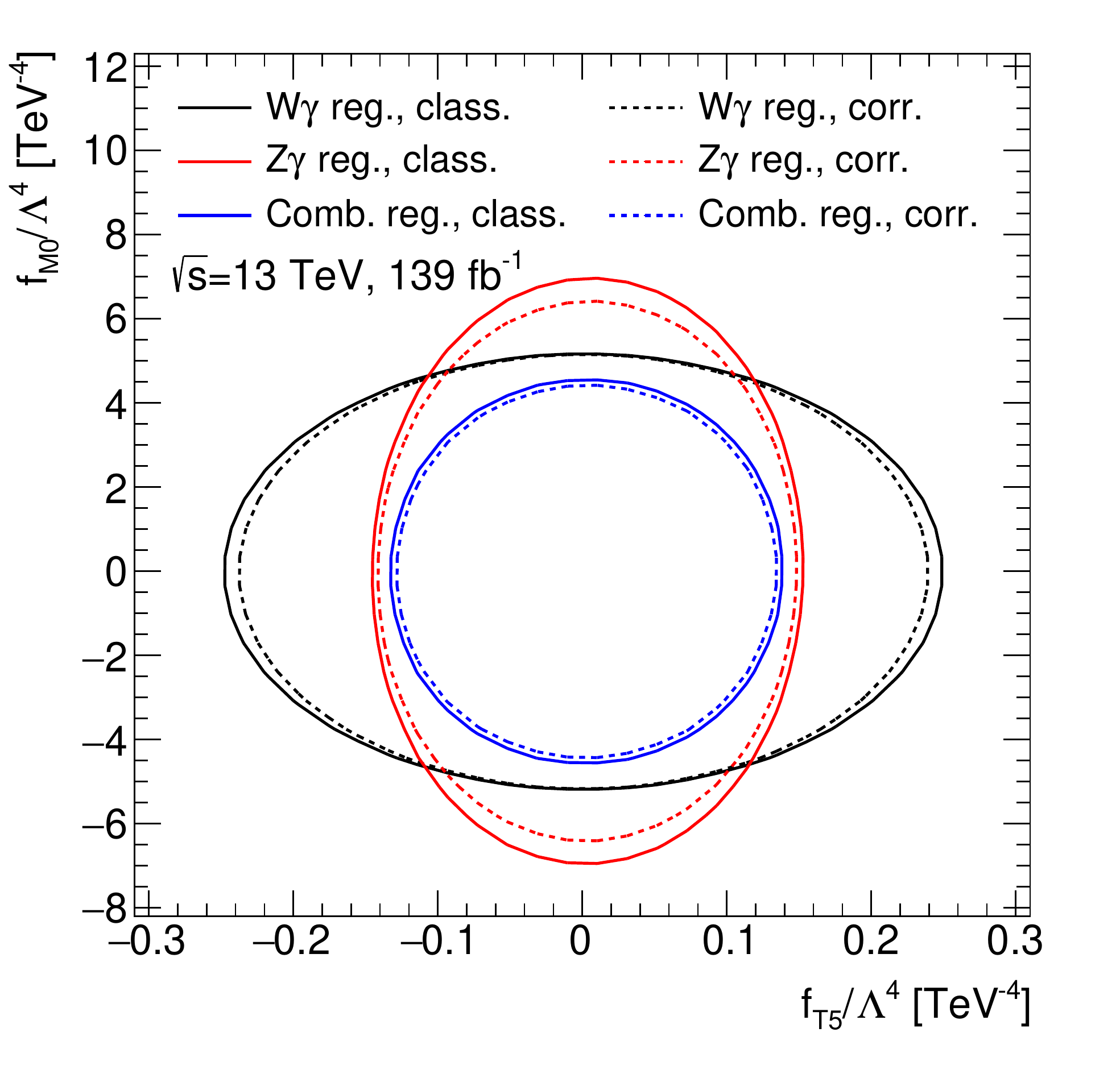}
    \end{minipage}
    \begin{minipage}{0.32\textwidth}
    \centering
    \includegraphics[width=0.95\textwidth]{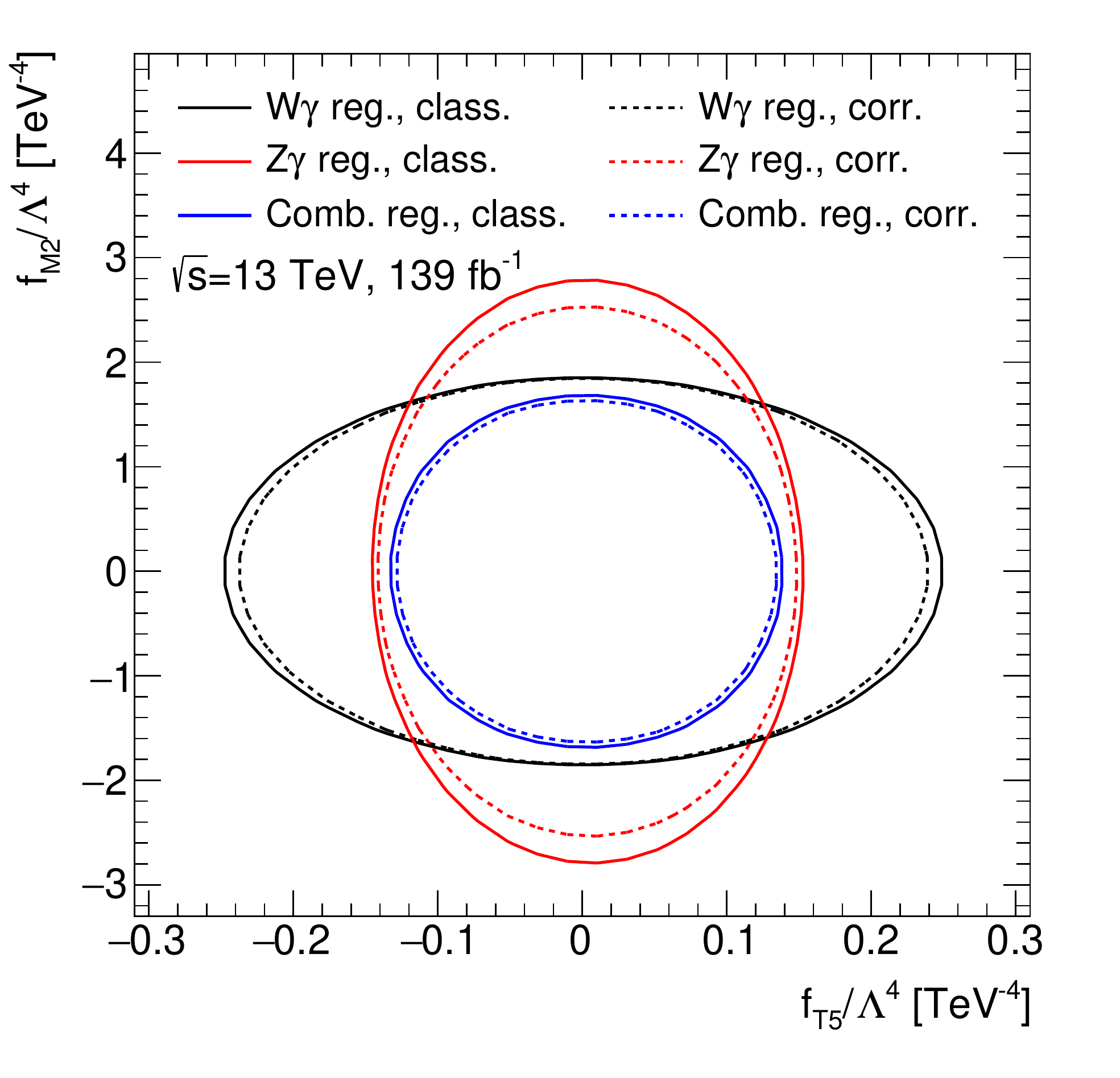}
    \end{minipage}
    \begin{minipage}{0.32\textwidth}
    \centering
    \includegraphics[width=0.95\textwidth]{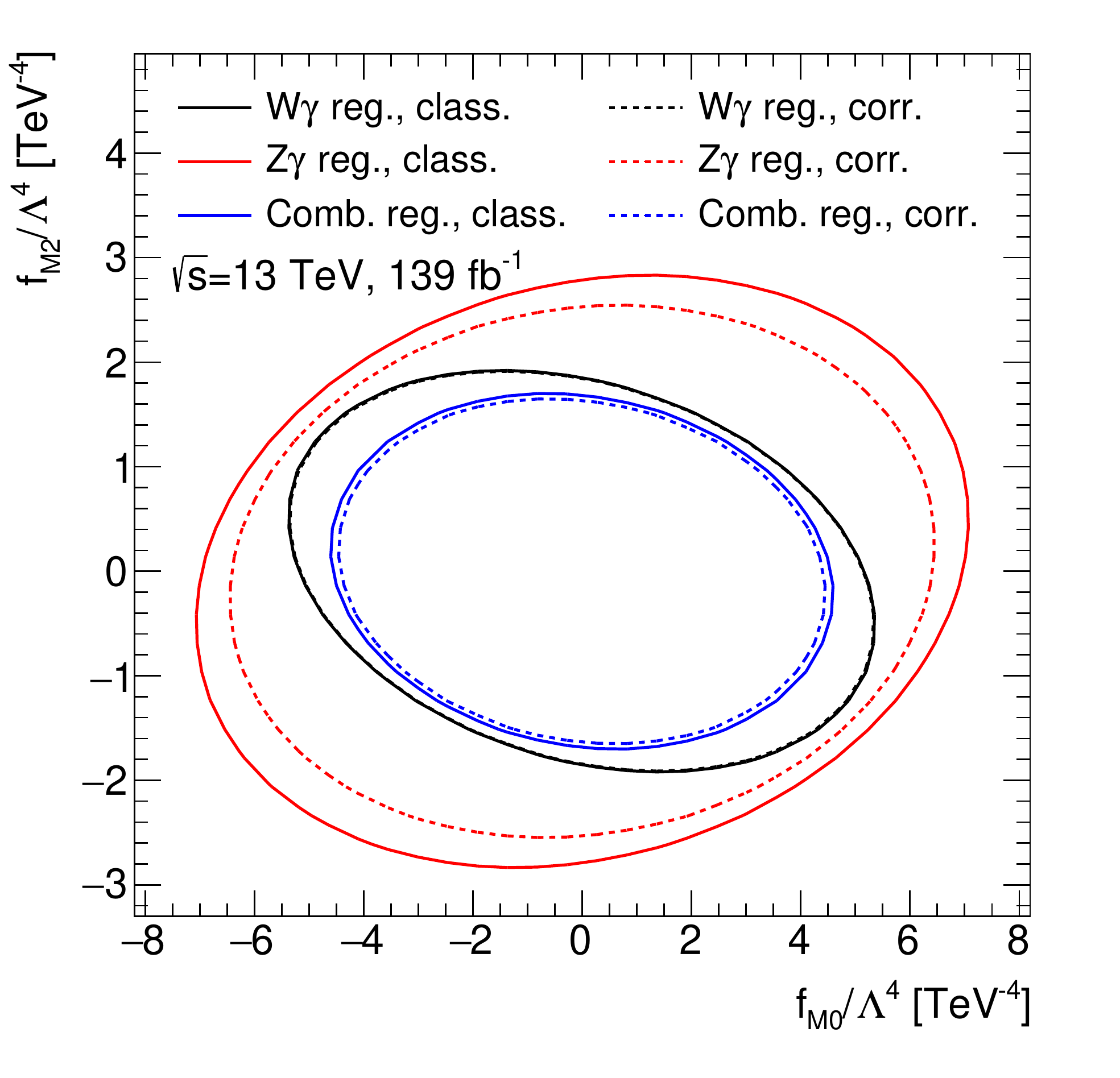}
    \end{minipage}
    \caption{The resulting classical (solid) and corrected (dashed) 2D limits for six combinations of Wilson coefficients, set from $Z\gamma$ region (red), $W\gamma$ region (black) and combined region (blue). The case of Run~II integrated luminosity.}
    \label{Results2D_139}
\end{figure}

\begin{figure}[h!]
    \centering
    \begin{minipage}{0.32\textwidth}
    \centering
    \includegraphics[width=0.95\textwidth]{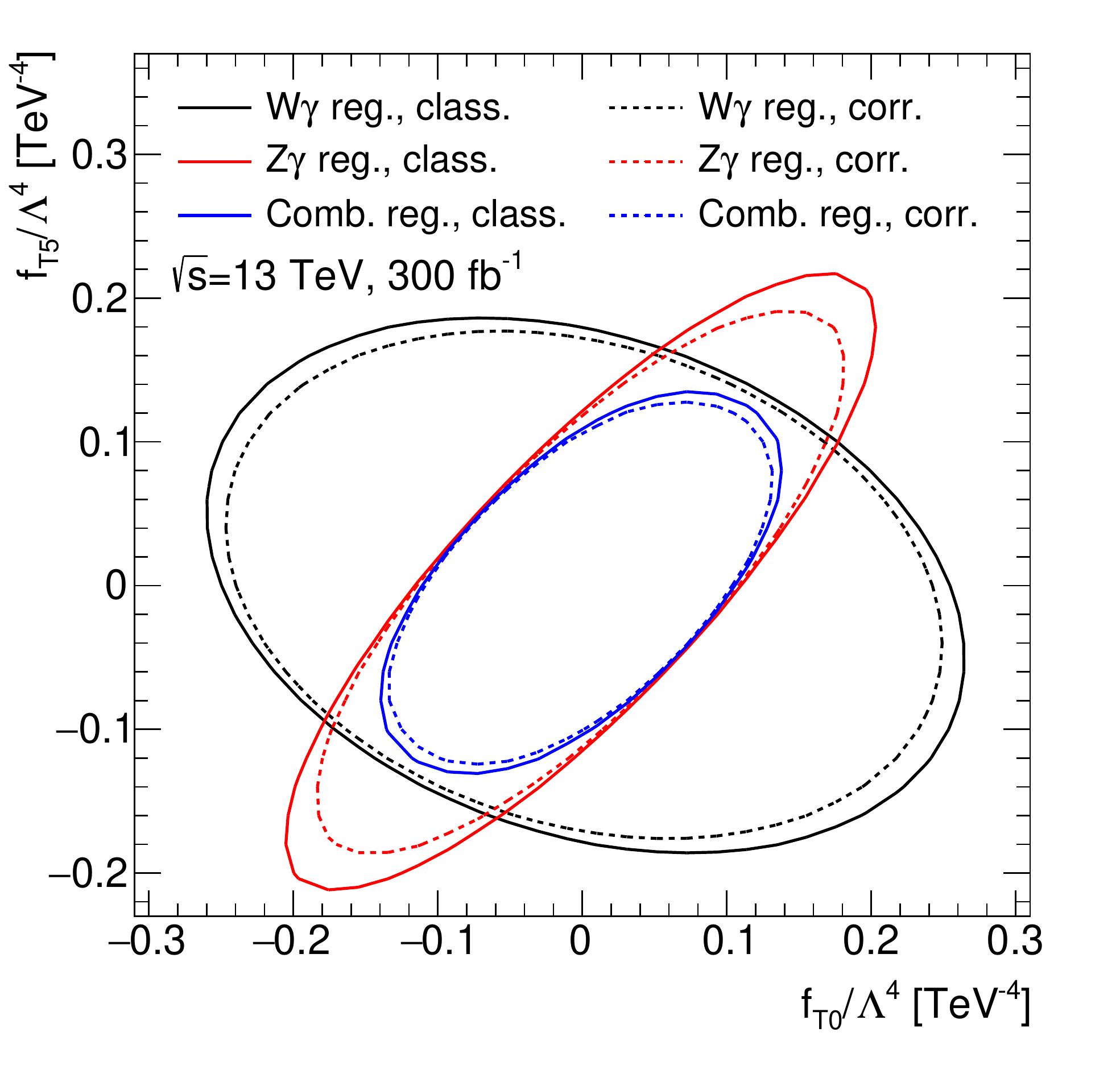}
    \end{minipage}
    \begin{minipage}{0.32\textwidth}
    \centering
    \includegraphics[width=0.95\textwidth]{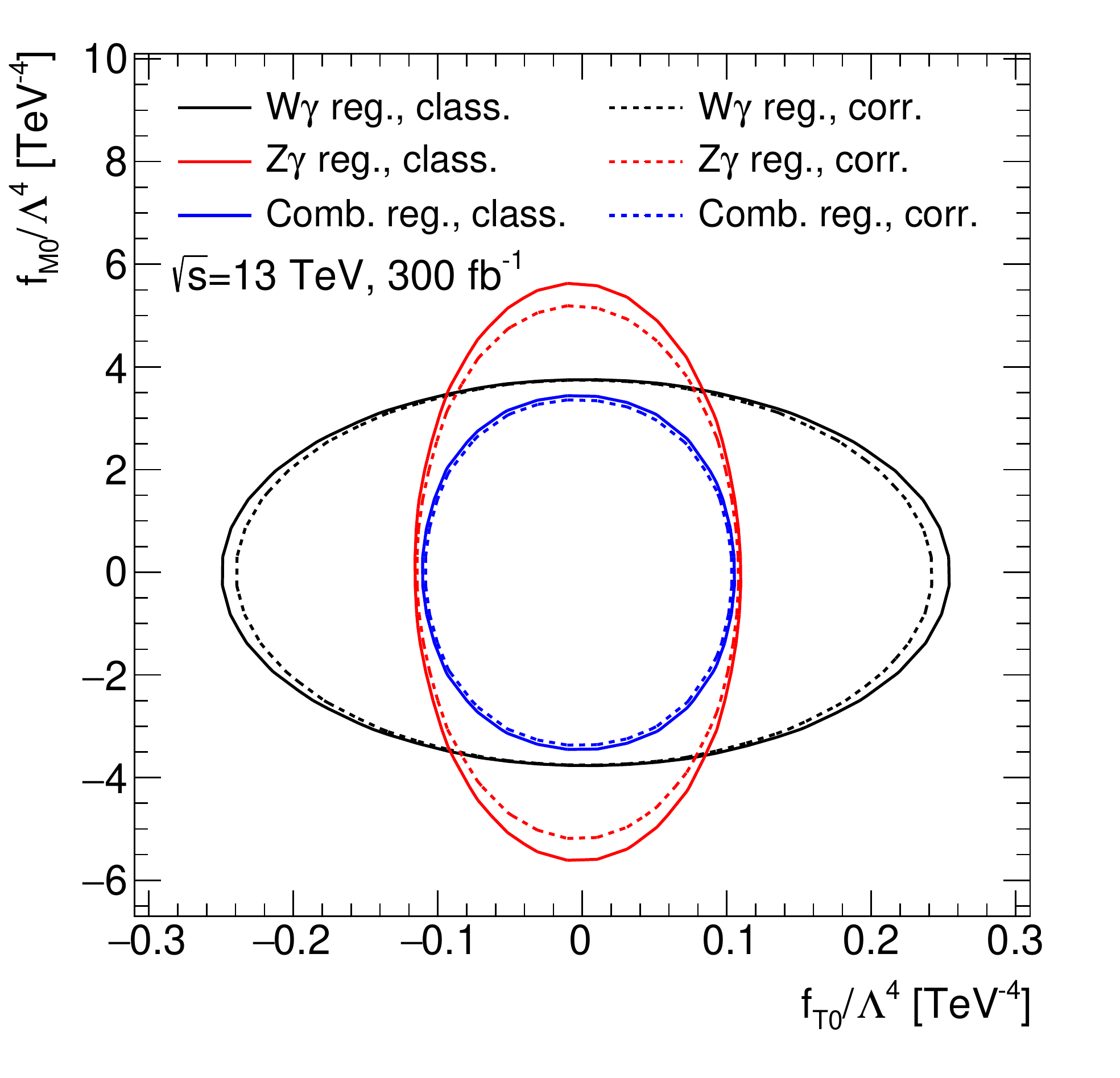}
    \end{minipage}
    \begin{minipage}{0.32\textwidth}
    \centering
    \includegraphics[width=0.95\textwidth]{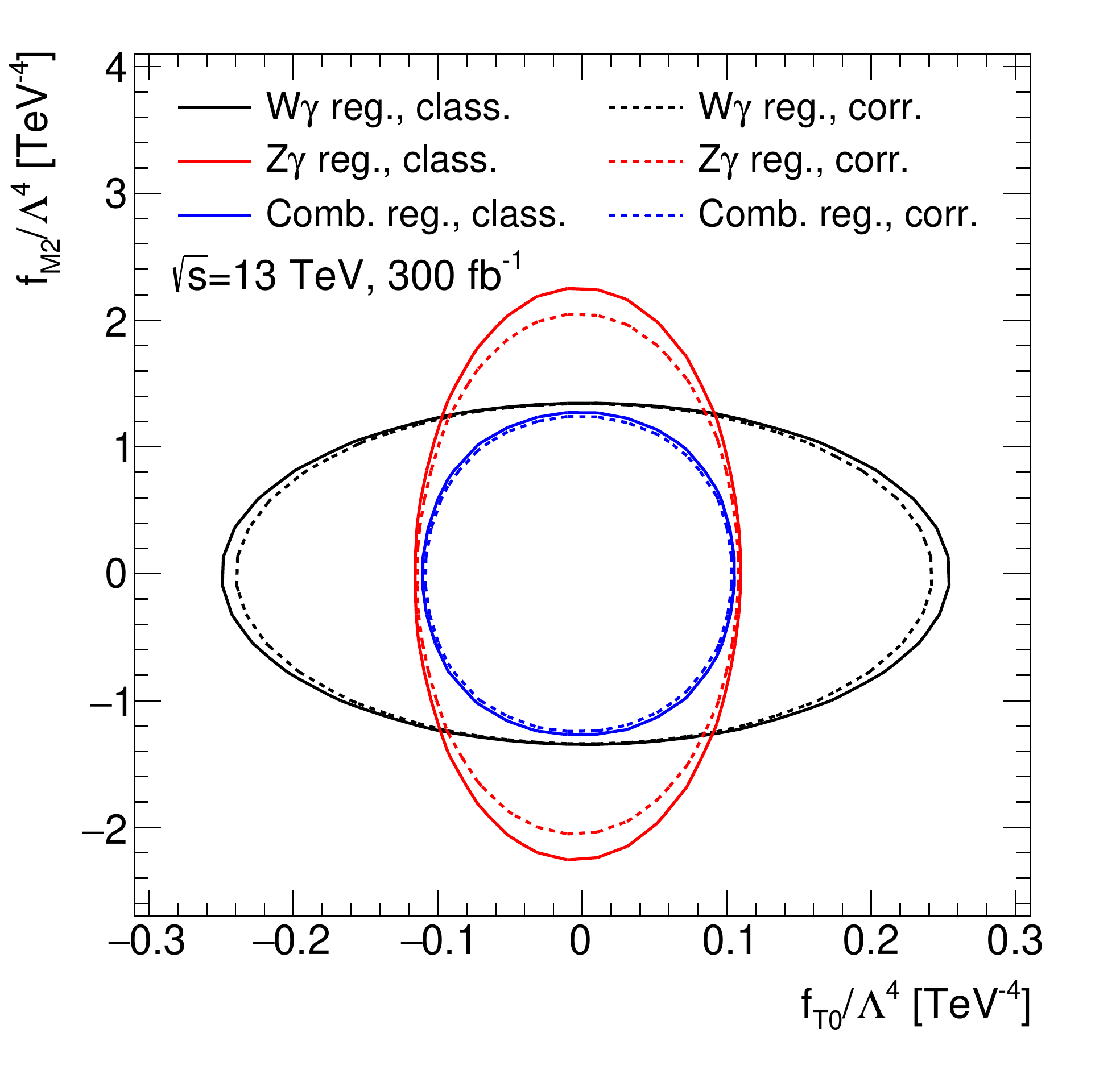}
    \end{minipage}
    \par
    \begin{minipage}{0.32\textwidth}
    \centering
    \includegraphics[width=0.95\textwidth]{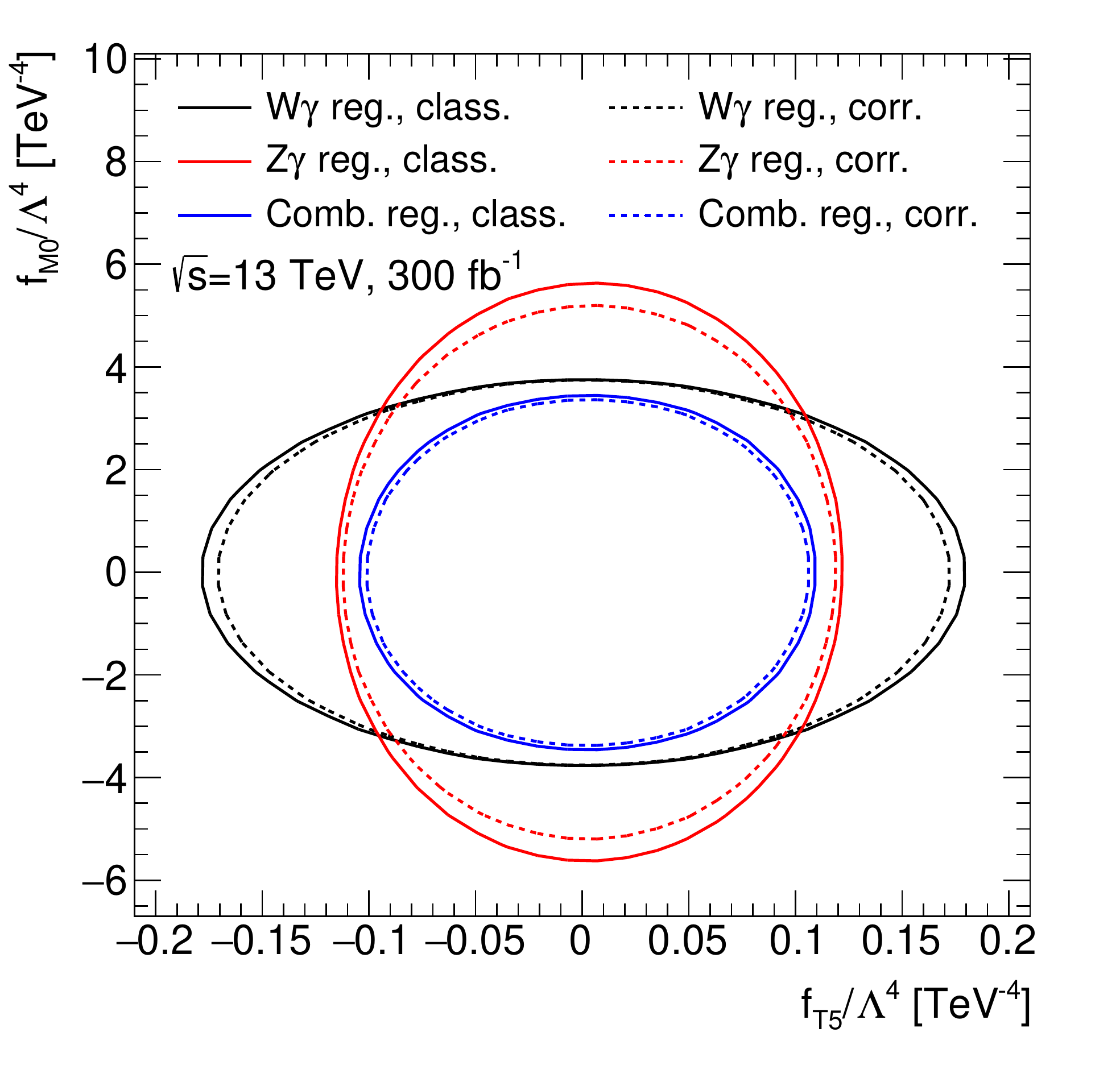}
    \end{minipage}
    \begin{minipage}{0.32\textwidth}
    \centering
    \includegraphics[width=0.95\textwidth]{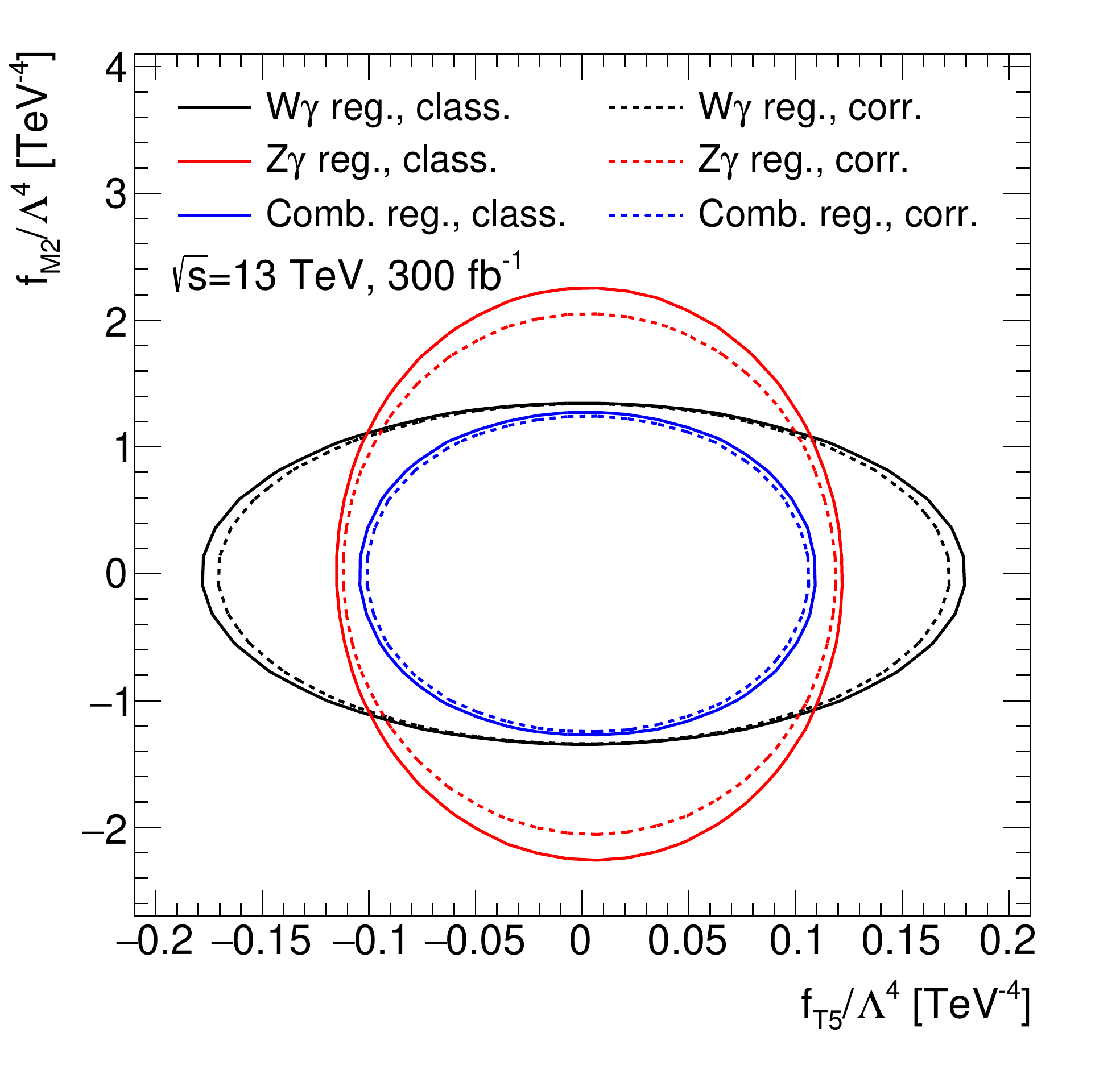}
    \end{minipage}
    \begin{minipage}{0.32\textwidth}
    \centering
    \includegraphics[width=0.95\textwidth]{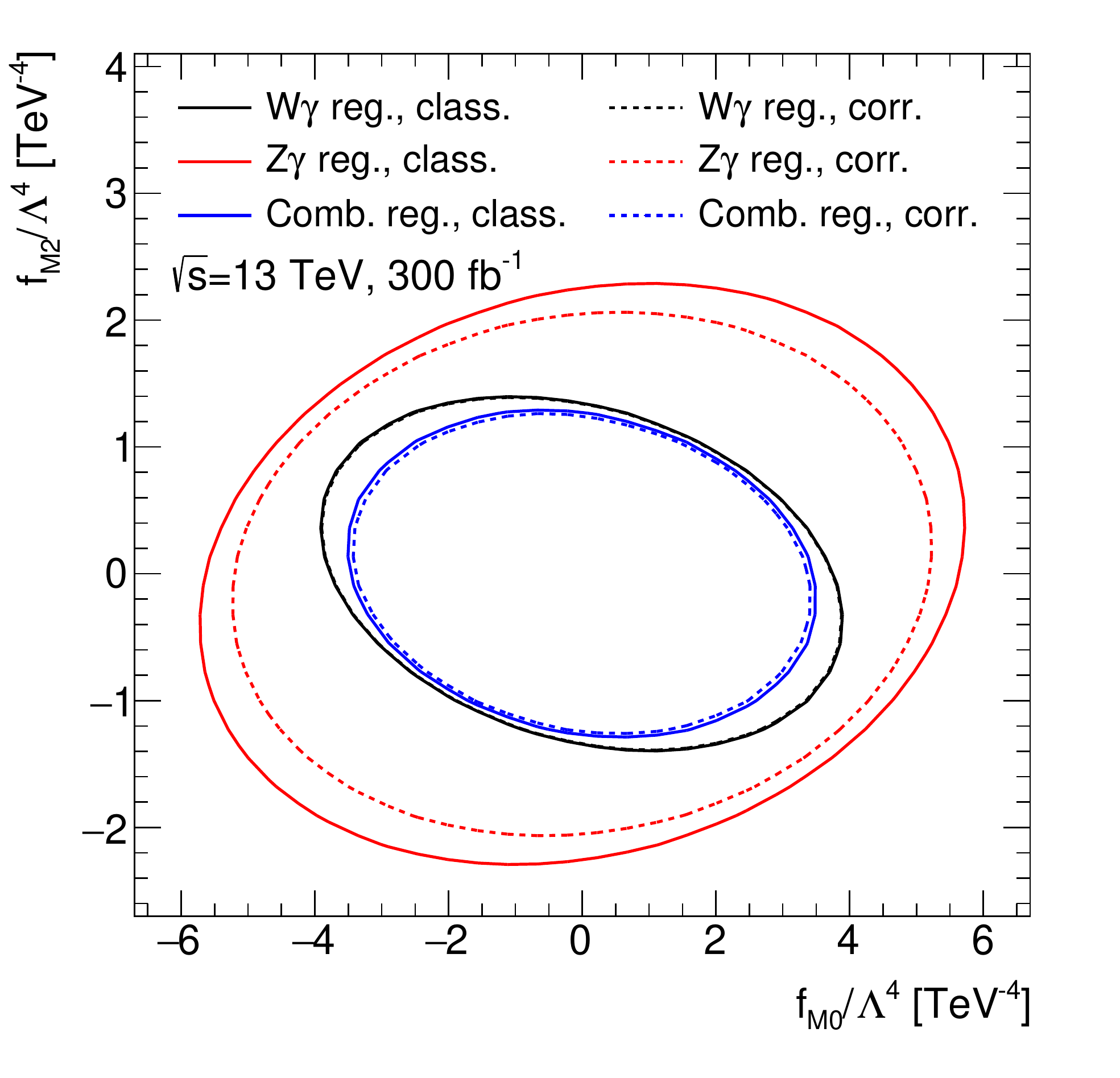}
    \end{minipage}
    \caption{The resulting classical (solid) and corrected (dashed) 2D limits for six combinations of Wilson coefficients, set from $Z\gamma$ region (red), $W\gamma$ region (black) and combined region (blue). The case of Run~III integrated luminosity.}
    \label{Results2D_300}
\end{figure}

The results for 2D 95\% CL expected limits (contours) for all six combinations of considered coefficients are presented in Figures~\ref{Results2D_139} and~\ref{Results2D_300} for integrated luminosity values of 139~fb$^{-1}$ and 300~fb$^{-1}$, respectively. The most stringent limits correspond to the combined region. All corrected limits (dashed) are more stringent compared with the classical ones (solid). The maximum improvement is in the $Z\gamma$ region for $\fmzero$ vs. $\fmtwo$ contour and is  17.2\% for 139~fb$^{-1}$ and 16.8\% for 300~fb$^{-1}$.

\newpage
\begin{table}[h!]
    \centering
    \caption{The resulting 1D limits on Wilson coefficients derived using the $Z\gamma$ region.}
    \label{Zgamma1DResults}
    \begin{tabular}{cccc}
        \toprule
        Coefficient & Class. lim. [TeV$^{-4}$] & Corr. lim. [TeV$^{-4}$] & Improvement \\ \midrule
        \multicolumn{4}{c}{Integrated luminosity of 139 fb$^{-1}$} \\ \midrule
        $\ftzero$ & [-0.125; 0.119] & [-0.124; 0.118] & 1.3\% \\
        $\ftfive$ & [-0.125; 0.132] & [-0.122; 0.128] & 2.7\% \\
        $\fmzero$ & [-6.04; 6.05] & [-5.57; 5.57] & 7.8\% \\
        $\fmtwo$ & [-2.42; 2.42] & [-2.20; 2.20] & 9.1\% \\ \midrule
        \multicolumn{4}{c}{Integrated luminosity of 300 fb$^{-1}$} \\ \midrule
        $\ftzero$ & [-0.102; 0.096] & [-0.100; 0.095] & 1.3\% \\
        $\ftfive$ & [-0.101; 0.107] & [-0.098; 0.104] & 2.6\% \\
        $\fmzero$ & [-4.95; 4.95] & [-4.57; 4.57] & 7.6\% \\
        $\fmtwo$ & [-1.98; 1.98] & [-1.81; 1.81] & 8.9\% \\ \bottomrule
    \end{tabular}
\end{table}
\begin{table}[h!]
    \centering
    \caption{The resulting 1D limits on Wilson coefficients derived using the $W\gamma$ region.}
    \label{Wgamma1DResults}
    \begin{tabular}{cccc}
        \toprule
        Coefficient & Class. lim. [TeV$^{-4}$] & Corr. lim. [TeV$^{-4}$] & Improvement \\ \midrule
        \multicolumn{4}{c}{Integrated luminosity of 139 fb$^{-1}$} \\ \midrule
        $\ftzero$ & [-0.286; 0.292] & [-0.274; 0.278] & 4.4\% \\
        $\ftfive$ & [-0.204; 0.206] & [-0.196; 0.198] & 4.1\% \\
        $\fmzero$ & [-4.30; 4.28] & [-4.29; 4.27] & 0.3\% \\
        $\fmtwo$ & [-1.54; 1.54] & [-1.53; 1.53] & 0.4\% \\ \midrule
        \multicolumn{4}{c}{Integrated luminosity of 300 fb$^{-1}$} \\ \midrule
        $\ftzero$ & [-0.207; 0.211] & [-0.199; 0.201] & 4.4\% \\
        $\ftfive$ & [-0.148; 0.149] & [-0.142; 0.143] & 4.1\% \\
        $\fmzero$ & [-3.14; 3.13] & [-3.14; 3.12] & 0.3\% \\
        $\fmtwo$ & [-1.13; 1.12] & [-1.12; 1.12] & 0.4\% \\ \bottomrule
    \end{tabular}
\end{table}
\begin{table}[h!]
    \centering
    \caption{The resulting 1D limits on Wilson coefficients derived using the combined region.}
    \label{Combination1DResults}
    \begin{tabular}{cccc}
        \toprule
        Coefficient & Class. lim. [TeV$^{-4}$] & Corr. lim. [TeV$^{-4}$] & Improvement \\ \midrule
        \multicolumn{4}{c}{Integrated luminosity of 139 fb$^{-1}$} \\ \midrule
        $\ftzero$ & [-0.120; 0.115] & [-0.118; 0.113] & 1.6\% \\
        $\ftfive$ & [-0.114; 0.119] & [-0.110; 0.116] & 3.0\% \\
        $\fmzero$ & [-3.86; 3.85] & [-3.76; 3.75] & 2.7\% \\
        $\fmtwo$ & [-1.43; 1.42] & [-1.39; 1.39] & 2.6\% \\ \midrule
        \multicolumn{4}{c}{Integrated luminosity of 300 fb$^{-1}$} \\ \midrule
        $\ftzero$ & [-0.0972; 0.0926] & [-0.0956; 0.0910] & 1.6\% \\
        $\ftfive$ & [-0.0906; 0.0955] & [-0.0879; 0.0927] & 3.0\% \\
        $\fmzero$ & [-2.94; 2.93] & [-2.88; 2.87] & 2.0\% \\
        $\fmtwo$ & [-1.07; 1.07] & [-1.05; 1.05] & 1.9\% \\ \bottomrule
    \end{tabular}
\end{table}

\newpage

\newpage
\section{Conclusions}
We investigated the methodology of considering  possible new physics manifestations in the background processes for setting the limits on anomalous couplings  in this article. In the full EFT model (with all terms), this methodology can lead to tightening as well as weakening of the limits in general case. In the case of Run II and Run III sensitivity, the limits were improved from both sides. Improvements of the limits in the $Z\gamma$ and $W\gamma$ regions for Run~II (Run~III) integrated luminosity were 1.3--9.1\% (1.3--8.9\%) and 0.3--4.4\% (0.3--4.4\%) depending on the operator. A combination of limits from these two regions was also considered in this work. The improvement in the combined region for both Run~II and Run~III cases was 1.6--3.0\% depending on the operator. Corrections to the two-dimensional limits for all coefficient combinations were calculated, and all corrected contours were more stringent compared with the classical ones, up to 17.2\% (16.8\%) for Run~II (Run~III) integrated luminosity. Thus, this methodology provides a possibility to make the limits on Wilson coefficients more correct and tight. Corrections for certain operators are sizeable.

In the same way, it is possible to correct aQGC limits obtained into the studies of production of other multiboson states as well as the limits on other types of anomalous couplings. For example, this methodology can be applied to the limits on anomalous triple gauge couplings. In this case, the corrections will be even more significant than the ones presented in this work, since there are more backgrounds that are sensitive to anomalous triple gauge couplings.

\section*{Acknowledgements}
This work was supported by the Russian Science Foundation under grant 21-72-10113.
\bibliographystyle{JHEP}
\bibliography{references1}

\end{document}